\documentclass[preprint2]{aastex}
\usepackage{amsmath,amssymb,amstext}
\usepackage{longtable}
\newcommand{\GALEX}{\emph{ GALEX} }
\usepackage[breaklinks,colorlinks,citecolor=blue,linkcolor=magenta]{hyperref} 
\usepackage[all]{hypcap} 
\usepackage{aas_macros}
\usepackage{natbib}
\usepackage{textcomp}
\usepackage{subfigure}
\usepackage{graphicx}
\usepackage{float}

\begin{document}

\title{HAZMAT II: Ultraviolet Variability of Low-Mass Stars in the GALEX Archive }
\author{Brittany E. Miles\altaffilmark{1,2,3,4} \\ \texttt{bmiles@ucsc.edu} \and  Evgenya L. Shkolnik\altaffilmark{5}}
        
\altaffiltext{1}{Department of Astronomy and Astrophysics, University of California, Santa Cruz}
\altaffiltext{2}{Department of Physics and Astronomy, University of California, Los Angeles}
\altaffiltext{3}{Lowell Observatory}
\altaffiltext{4}{Department of Physics and Astronomy, Northern Arizona University}
\altaffiltext{5}{School of Earth and Space Exploration, Arizona State University}

\begin{abstract}

The ultraviolet (UV) light from a host star influences a planet's atmospheric photochemistry and will affect interpretations of exoplanetary spectra from future missions like the James Webb Space Telescope. These effects will be particularly critical in the study of planetary atmospheres around M dwarfs, including Earth-sized planets in the habitable zone. Given the higher activity levels of M dwarfs compared to Sun-like stars, time resolved UV data are needed for more accurate input conditions for exoplanet atmospheric modeling. The Galaxy Evolution Explorer (\emph{GALEX}) provides multi-epoch photometric observations in two UV bands: near-ultraviolet (NUV; 1771 -- 2831 \AA) and far-ultraviolet (FUV; 1344 -- 1786 \AA). Within 30 pc of Earth, there are 357 and 303 M dwarfs in the NUV and FUV bands, respectively, with multiple\GALEX observations. Simultaneous NUV and FUV detections exist for 145 stars in both\GALEX bands. Our analyses of these data show that low-mass stars are typically more variable in the FUV than the NUV. Median variability increases with later spectral types in the NUV with no clear trend in the FUV. We find evidence that flares increase the FUV flux density far more than the NUV flux density, leading to variable FUV to NUV flux density ratios in the \GALEX\ bandpasses.The ratio of FUV to NUV flux is important for interpreting the presence of atmospheric molecules in planetary atmospheres such as oxygen and methane as a high FUV to NUV ratio may cause false-positive biosignature detections. This ratio of flux density in the\GALEX\ bands spans three orders of magnitude in our sample, from 0.008 to 4.6, and is 1 to 2 orders of magnitude higher than for G dwarfs like the Sun. These results characterize the UV behavior for the largest set of low-mass stars to date. 
\\
\end{abstract}
\maketitle

\section{Introduction}
 M dwarfs provide excellent laboratories for understanding the diversity of exoplanets  as they represent 75\% of the stars in the Milky Way \citep{2010AJ....139.2679B} and are prime targets for finding small, habitable zone (HZ), rocky exoplanets. An Earth-sized planet in the HZ around an M-type star produces deeper transits and has a shorter follow-up time compared to a HZ planet orbiting a G type star. Using four years of Kepler data, modeling, and follow-up observations, \citet{2015ApJ...807...45D} estimated that on average there are at least two small planets around every M dwarf, with at least one Earth-size HZ planet for every seven M dwarfs. In fact, a $>$1.3 M$_{\Earth}$ planet has been found orbiting in the HZ of our nearest neighbor, mid-M dwarf Proxima Centauri \citep{2016Natur.536..437A} and three Earth-sized HZ planets have been found around the late-M star TRAPPIST-1 \citep{2017Natur.542..456G}.   
 
 Despite all this, M dwarfs come with complications to characterizing exoplanets. For example, the deep convective zones of low-mass stars create non-uniform magnetic fields that rise above the stellar surface, expelling a large amount of energy in the chromosphere and corona. This energy bombards exoplanets with short-wavelength photons and high-energy particles from stellar winds. With such exposure to a variable and energetic environment, one cannot characterize an exoplanet's atmosphere and habitability without considering the full impact of the host star.
 
 Several teams are studying the atmospheric response of planets exposed to strong emission and variability in the UV and X-ray as they are the most damaging wavelengths to atmospheric photochemistry (e.g.~\citealt{2014ApJ...780..166M,2015ApJ...809...57R,2015AsBio..15...57L, 2015AsBio..15..119L,arne16}).  Models that use real UV spectra (e.g.~\citealt{2013ApJ...763..149F}) show that detecting and understanding the sources of oxygen gas and other molecules requires knowing the depletion rate and ozone (O$_{3}$) build-up caused by the high-energy radiation from the host star. This is especially true when molecular lines are being used to infer biological processes \citep{2015ApJ...809...57R,2015AsBio..15..119L}.
 
 In an extreme case, \citet{2010AsBio..10..751S} investigated the effect of the strongest UV flare (10$^{33}$ - 10$^{34}$  ergs) observed from the very active M dwarf, AD Leo, on the atmosphere of an Earth-like planet in the HZ. They observed  a several orders of magnitude decrease in upper-atmosphere O$_{3}$ and water abundances within a day with smaller scale fluctuations leading to equilibrium over the course of a year following the flaring event. Even for  stars deemed ``inactive'', due to the lack of H$\alpha$ emission, time-tagged Hubble Space Telescope (\textit{HST})  UV spectra from four M dwarf planet hosts displayed changes in emission lines ranging from 50 - 500\% on the order of minutes \citep{2013ApJ...763..149F, 2014ApJS..211....9L}.

For stars that are K7 or later, the published relationships between contemporaneous UV bands and variability are limited to samples of less than 10 stars \citep{2005A&A...431..679M,2016ApJ...820...89F}. Monitoring programs for a large sample of M dwarfs are challenging to execute with \textit{HST} and there are limited resources for studying these objects at very short wavelengths. Measuring the UV flux from M dwarfs needs to be extended to a significantly larger sample of stars to understand the full range of emission levels and variability. 

The space-based telescope Galaxy Evolution Explorer (\textit{GALEX}, \citealt{2005ApJ...619L...7M}) provides the opportunity to study a much larger sample of M dwarfs than has been possible before. Operational from 2003 to 2012,\GALEX tiled over 2/3 of the sky with two UV filters, often simultaneously, with 1.2$^{\circ}$-diameter images, capturing UV data for hundreds of low-mass stars. The near-ultraviolet  (NUV; 1771 -- 2831 \AA) and far-ultraviolet (FUV; 1344 -- 1786 \AA) bands are excellent probes for high energy stellar activity because they are composed of emission lines formed in the chromosphere and corona of stars \citep{2009AcGeo..57...42K, 2006A&A...458..921W}. Previous variability studies have been conducted with\GALEX, but focus on a small set of low-mass stars \citep{2008AJ....136..259W}. In this second paper of \textbf{HA}bitable \textbf{Z}ones and \textbf{M} dwarf \textbf{A}ctivity across \textbf{T}ime (HAZMAT) series, we use archived data from both \GALEX\ photometric bands to measure the variability of 376 low-mass stars with spectral types ranging from K7 to M7, and analyze the relationship between the NUV and FUV emission for simultaneous and time-resolved observations.

\section{Low-Mass Star Sample and\GALEX Data}
\label{sec:Star Sample}

Our target list consisted of 1124 low-mass stars with photometric distances out to 25 pc of Earth assuming field ages  \citep{2007AJ....133.2825R}. Once accounting for the young stars in the sample whose distances are in fact slightly further, all targets are within 30 pc. Ages were primarily taken from the HAZMAT I paper,  \cite{2014AJ....148...64S} and \cite{2013MNRAS.431.2063S} and when available parallactic distances were compiled from \cite{2009ApJ...699..649S}, \cite{2012ApJ...758...56S} and the Hipparcos catalogue \citep{1997A&A...323L..49P}. 

Most of the spectral type identifications are from \cite{2007AJ....133.2825R} and these identifications were then confirmed with a literature search. If no spectral identification was available from the original list, a literature measurement is taken with a preference for optical identifications. There are sometimes discrepancies between published stellar types, therefore measurements from large surveys such as the Palomar/MSU survey (\citealt{1995AJ....110.1838R}, \cite{1996AJ....112.2799H}) and Meeting the Cool Neighbors Series \citep{2002AJ....123.2806R} are primarily used for consistency. All references for spectral types are listed in Table~\ref{tbl:starinfo}. The seven stars with an ``M:'' spectral type designation either do not have a literature measurement or have discrepant published values with a $>1$ spectral type subclass difference. These stars are excluded from the analysis regarding spectral type and \GALEX band correlations.

\begin{figure}[ht!]
\centering
\includegraphics[width = 3in]{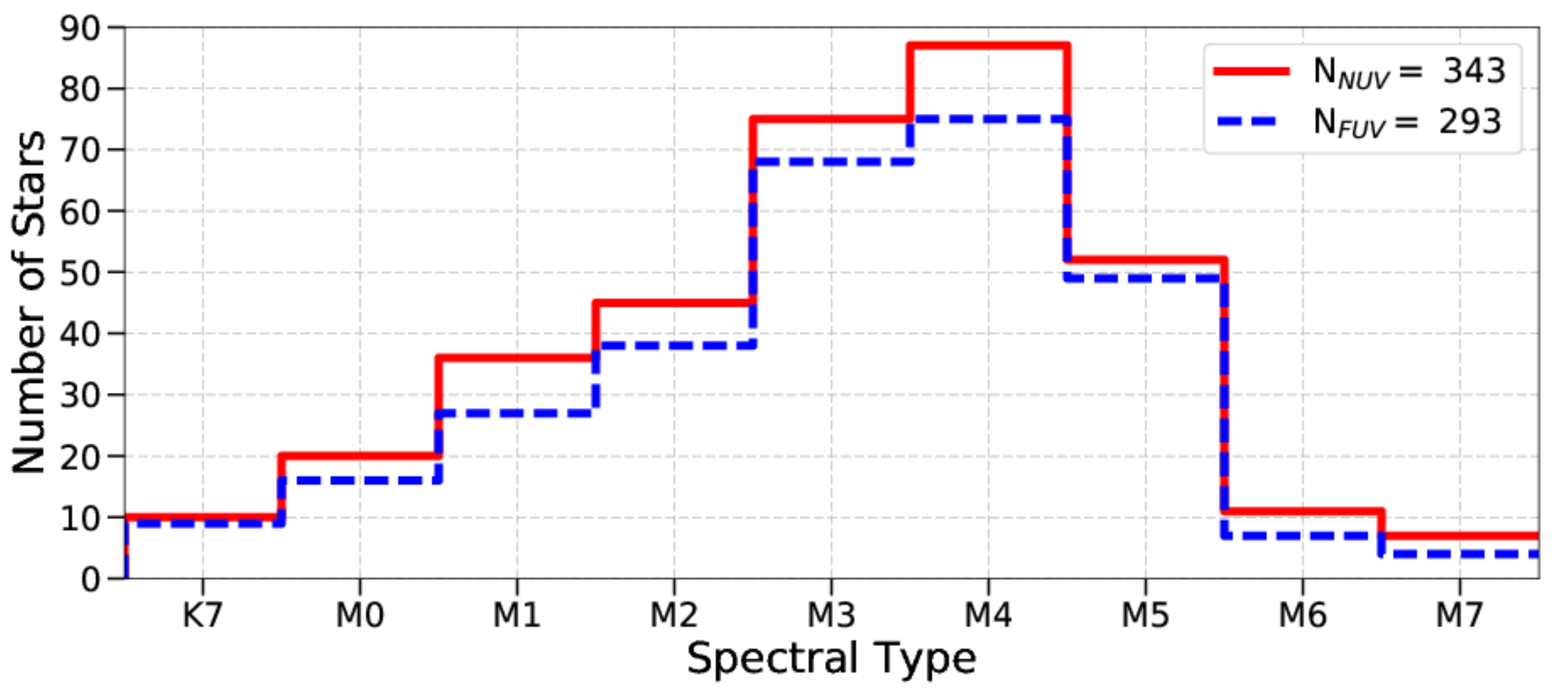}
\caption{The distribution of spectral types for stars with multiple NUV (red) and FUV (blue) observations including $\geq$2$\sigma$ detections and upper limits.}
\label{fig:spthist} 
\end{figure}

\newpage
\begin{deluxetable}{lccccccccccccccccc}
\tabletypesize{\tiny} 
\rotate
\tablecaption{Low-Mass Stars with Multiple \GALEX UV Observations\label{tbl:starinfo}}
\tablewidth{0pt} 
\tablecolumns{15}

\tablehead{
\colhead{Name\tablenotemark{a} } & 
\colhead{RA} & 
\colhead{DEC} & 
\colhead{SpT} &
\colhead{Age\tablenotemark{b} } &
\colhead{Ref } &
\colhead{Dist.} &
\colhead{NUV} &
\colhead{NUV} &
\colhead{NUV}  &
\colhead{NUV} &
\colhead{FUV} &
\colhead{FUV} &
\colhead{FUV} &
\colhead{FUV}\\ 

\colhead{} & 
\colhead{J2007} & 
\colhead{J2007} & 
\colhead{} &
\colhead{} &
\colhead{} &
\colhead{} &
\colhead{det\tablenotemark{c},u.l.\tablenotemark{d} } &
\colhead{$\mu$\tablenotemark{e}} &
\colhead{Min,Max\tablenotemark{g} } &
\colhead{MAD} &
\colhead{det\tablenotemark{c},u.l.\tablenotemark{d}} &
\colhead{$\mu$\tablenotemark{e}} &
\colhead{Min,Max\tablenotemark{g}} & 
\colhead{MAD}\\

\colhead{} & 
\colhead{(deg)} & 
\colhead{(deg)} & 
\colhead{} &
\colhead{(Myr)} &
\colhead{} &
\colhead{(pc)} &
\colhead{} &
\colhead{($\mu$Jy)} &
\colhead{($\mu$Jy),($\mu$Jy)} &
\colhead{rel\tablenotemark{f}  } &
\colhead{} &
\colhead{($\mu$Jy)} &
\colhead{($\mu$Jy), ($\mu$Jy)}  &
\colhead{ rel\tablenotemark{g}}\\ 
}

\startdata

LHS1014 & 1.29780 & +45.78628 & M1 & 5000 & -,1 & 11.5 & 2,- & 158.26$\pm$9.16 & 145.40,171.13 & 0.08 & -,- & - & - & -\\
NLTT400 & 2.30600 & -4.13416 & M2.5 & 5000 & -,12 & 23.8 & 6,- & 78.76$\pm$5.15 & 69.28,94.54 & 0.07 & 2,3 & 24.61$\pm$1.23 & 7.76,17.84 & 0.56\\
NLTT557 & 2.97127 & +22.98422 & M3.5 & 2000 & 14,1 & 16.0 & 3,- & 16.43$\pm$3.45 & 12.34,22.89 & 0.12 & -,2 & $\leq$ 39.21 & - & 0.54\\
00144767-6003477 & 3.69901 & -60.06337 & M3.5 & 40 & 7,7 & 25.1 & 2,- & 438.25$\pm$25.23 & 298.25,578.29 & 0.32 & 2,- & 90.60$\pm$18.72 & 62.69,118.50 & 0.31\\
GJ1005 & 3.86861 & -16.13497 & M4 & 5000 & -,1 & 5.0 & 2,- & 1.74$\pm$0.43 & 1.01,2.47 & 0.42 & -,2 & $\leq$ 1.49 & - & 0.10\\
GJ3017 & 3.90328 & -29.76701 & M4 & 5000 & -,2 & 18.1 & 2,- & 87.22$\pm$10.26 & 62.57,111.85 & 0.28 & 1,1 & 22.06$\pm$4.13 & 0.00,0.00 & 0.20\\
LHS1051 & 3.96802 & -67.99805 & M0.5 & 5000 & -,2 & 19.8 & 2,- & 153.51$\pm$4.46 & 144.19,162.85 & 0.06 & -,- & - & - & -\\
GJ3030 & 5.49148 & +49.21050 & M2.4 & 150 & 13,13 & 24.8 & 2,- & 400.83$\pm$27.59 & 326.46,475.24 & 0.19 & 1,1 & 81.64$\pm$14.49 & 43.48,119.81 & 0.47\\
\enddata
\vspace{-.3cm}

\tablecomments{In the reference column, the first number is the spectral type reference and the second is the age reference. Reference numbers: 1. \cite{2007AJ....133.2825R} 2. \cite{1995AJ....110.1838R}, \cite{1996AJ....112.2799H}, 3.\cite{2015A&A...577A.128A} 4. \cite{2003AJ....126.2048G}, 5. \cite{2006AJ....132..161G}, 6. \cite{2009ApJ...704..975J}, 7. \cite{2014AJ....147..146K}, 8. \cite{2013AJ....145..102L} 9. \cite{2015ApJ...804...64M}, 10. \cite{2014AJ....147...20N}, 11. \cite{2013A&A...556A..15R}, 12. \cite{2005A&A...442..211S}, 13. \cite{2009ApJ...699..649S}, 14. \cite{2012ApJ...758...56S}, 15. \cite{2014ApJ...796L..20S}, 16. \cite{2006A&A...460..695T} 17. \cite{2012AJ....143...80S}.  The star GJ447 has several literature estimates for its age.  \cite{2001MNRAS.328...45M} determined that it is a member of the Ursa Major moving group (300 Myr) with a radial velocity (RV) of -13.0 $\pm$ 5.0 km/s. However, \cite{2003AJ....125.1980K} does not confirm the membership with a RV measurement of -29.0 $\pm$ 1.2  km/s, which is consistent with the RV standard list of \cite{2002ApJS..141..503N} (-31 $\pm$ 0.1 km/s).}
\tablenotetext{a}{2MASS designation}
\tablenotetext{b}{If no age is provided by the literature, the star is assumed to be 5 Gyr old.}
\tablenotetext{c}{det - detections}
\tablenotetext{d}{u.l. - upper limits}
\tablenotetext{e}{$\mu$ = Mean flux density values are scaled to 10 parsecs.}
\tablenotetext{f}{MAD rel- median absolute deviation (MAD) divided by the median indicated by MAD$_{rel}$ in our plots}
\tablenotetext{g}{Minimum and maximum flux density scaled to 10 parsecs}

\tablenotetext{h}{IRXS source, there is only one}

\end{deluxetable}

\subsection{\GALEX Photometry}
\label{sec:data}

The\GALEX pipeline performs photometry on all recognized sources within the field of view for every exposure \citep{2007ApJS..173..682M}.  We queried the archive\footnote{http://galex.stsci.edu/casjobs/} for all observations within a 5\arcsec\ radius centered around a target's proper motion corrected coordinates. If the cone search yielded multiple observations for a target at the same point in time, the observation closest to the coordinates were taken. Detections with a signal-to-noise of less than two were removed.\GALEX observed 55\% of the stars in our target list. In the NUV band, 36\% of the targets had at least two observations. In the FUV, 30\% were observed two or more times. Known spectroscopic binaries, visual binaries and background galaxies within the 5\arcsec \GALEX NUV point spread function were removed from our analysis. 

\begin{figure*}[ht!]
\centering
\subfigure{\includegraphics[width = 6in]{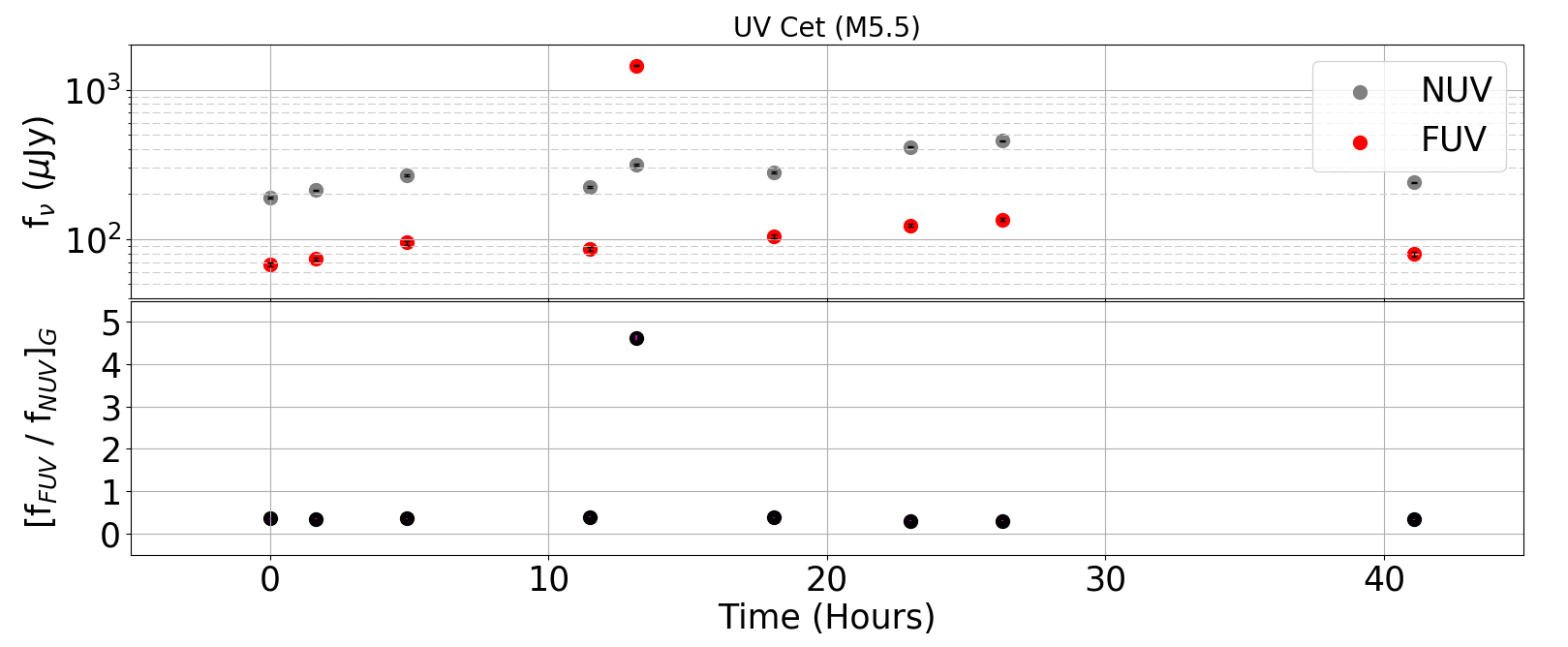}}
\subfigure{\includegraphics[width = 6in]{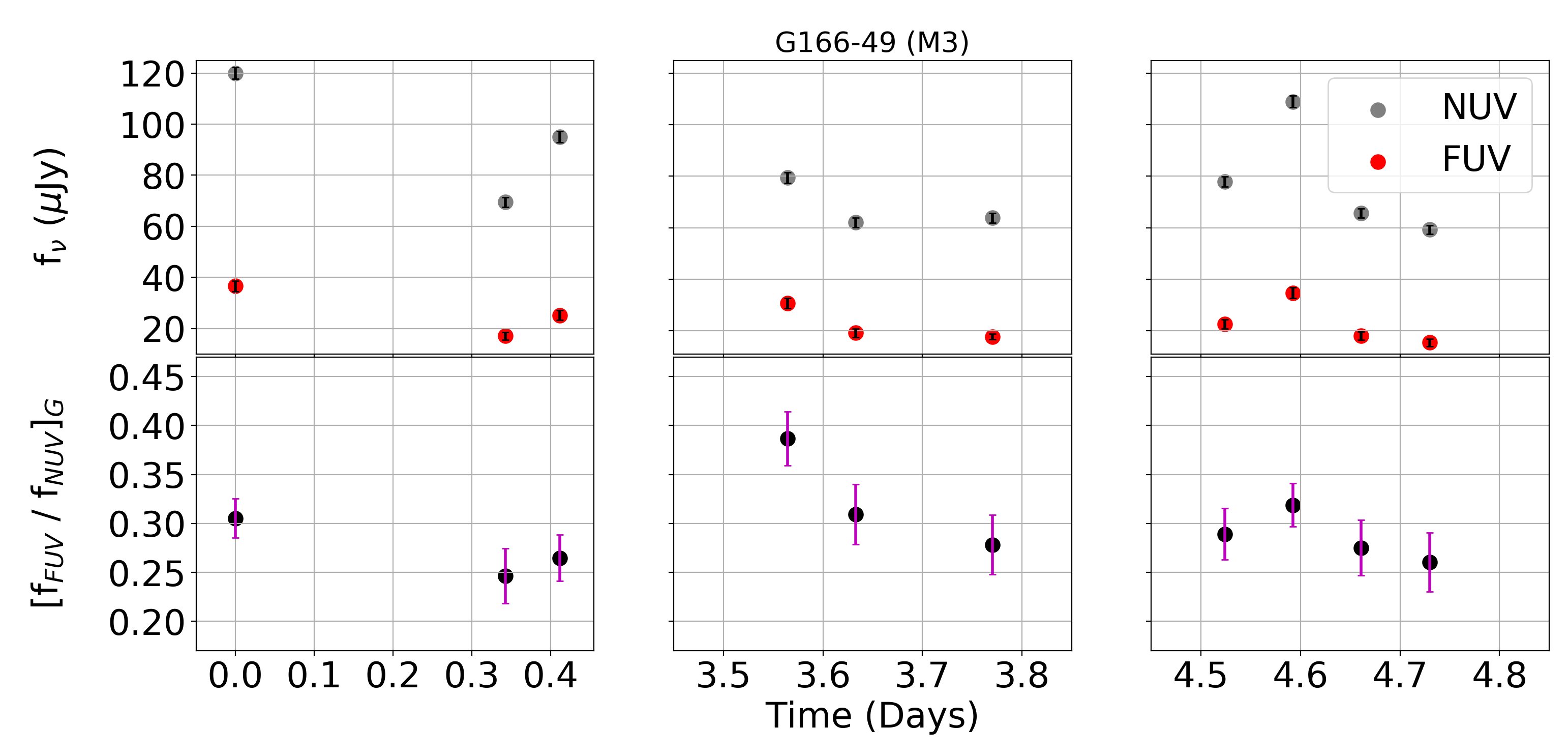}}
\caption{NUV (grey) and FUV (red) observed flux density light curves for UV Cet (top) and G166-49 (bottom). The ratio of FUV to NUV flux density ([f$_{FUV}$ /f$_{NUV}$]$_{G}$)is plotted with black dots below the light curves. The error bars are shown on the data points in black and magenta, for the flux densities and flux density ratios respectively. The units of time are different between both light curves. The NUV and FUV flux densities are typically correlated, but there is a strong flare on UV Cet where the FUV emission spikes compared to the NUV emission.}
\label{fig:lightcurves}
\end{figure*}

A non-linear response is recorded on the\GALEX detector when a count rate over 108 counts/s in the NUV or 34 counts/s in the FUV was reached \citep{2007ApJS..173..682M}. 2034 observations were taken in the NUV band. One NUV data point was removed for non-linearity and 54 were removed due to either reflections or ghosting on the NUV detector.\footnote{http://galex.stsci.edu/gr6/?page=ddfaq\\http://www.galex.caltech.edu/wiki/\\Public:Documentation/Chapter\_8\#Setting\_the\_Ghost\_Flag} There were 1403 FUV observations and only one point was removed for non-linearity. After these cuts, we were left with 357 and 303 stars with at least two reliable observations in the NUV and FUV band, respectively. At total of 1497 NUV and 1035 FUV observations were used in this analysis. Stars that are M5 and earlier represent 96\% of the stars with multiple detections in both photometric bands. None of the eight M8 or later stars from the original list have multiple\GALEX detections (Figure~\ref{fig:spthist}). Two examples of NUV and FUV light curves are shown in Figure~\ref{fig:lightcurves}.

The reported quantities from the \GALEX\ pipeline were used to estimate the upper limits, ensuring that all flux limits were extracted and calibrated in the same fashion. The 2$\sigma$ errors from nearby detections within 10$\arcmin$ and no\GALEX artifact flags can be used to estimate the upper limit of a non-detection. Upper limits were calculated for 0.9\% of the NUV and 59\% of the FUV observations for stars in our sample with multiple detections. The non-detection rate is 1\% higher in both bands when considering stars with at least one detection. With these detection rates, we achieved a volume limited sample in the NUV. The sample in the FUV is magnitude limited and we discuss the effect of this in Section~\ref{sec:fuvvar}.


\section{Variability Analysis}
\label{sec:analysis}

\begin{figure*}[ht!]
\centering
\includegraphics[width = 6in]{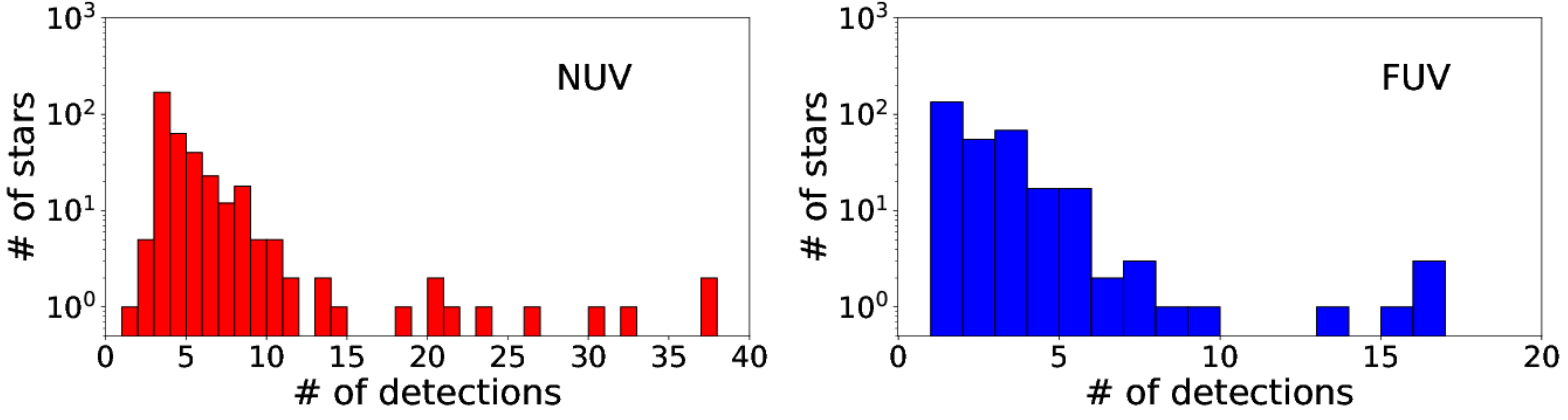}
\caption{Histogram of the number of detections for each\GALEX band. LHS 3776, which has 81 detections in the NUV was left off of the histogram.}
\label{fig:det_hist} 
\end{figure*}

The median number of detections for the NUV and FUV band is 3 and 1 respectively (Figure~\ref{fig:det_hist}).  This makes it difficult to characterize specific sources of a star's variability for each target such as rotation, flaring or changes in its activity cycle.  For those with at least two observations, the time span between observations ranges from two minutes to the full eight year mission. In order to systematically study the variability of the sample we computed the median absolute deviation divided by the median (MAD$_{rel}$) for the NUV and FUV flux densities for each target. We also report the maximum and minimum flux densities for each star in Table~\ref{tbl:starinfo}. Note that the minimum flux density may be from a detection or an upper limit.

The\GALEX pipeline calibrates flux densities and magnitudes to 18 white dwarf standards \citep{2007ApJS..173..682M}, which are assumed to be invariant and ideal UV flux calibrators \citep{1988ESASP.281b.365H}. However, we still measure variability in the light curves of the 18 \GALEX standard white dwarfs and the white dwarfs from the Villanova Catalog of Spectroscopically Identified White Dwarfs \citep{1999ApJS..121....1M}. We calculated the MAD$_{rel}$  of 1225 white dwarfs in the NUV and 952 white dwarfs in the FUV to cover a wide range of\GALEX magnitudes.\footnote{Pulsating white dwarfs, binary systems, and suspected planetary nebula were removed from the sample.}  We measured MAD$_{rel}$ to be 2.1\% and 3.5\%  for the NUV and FUV band, respectively. These photometric errors were added to the MAD$_{rel}$ uncertainties for all target stars in our analysis.
 
A summary of the variability results for individual targets with multiple detections are listed in Table~\ref{tbl:starinfo}. The results binned by spectral type are listed in Table~\ref{tbl:activtysummary}

\subsection{NUV Variability}

\begin{figure*}[h!]
\subfigure{\includegraphics[width = 6.5in]{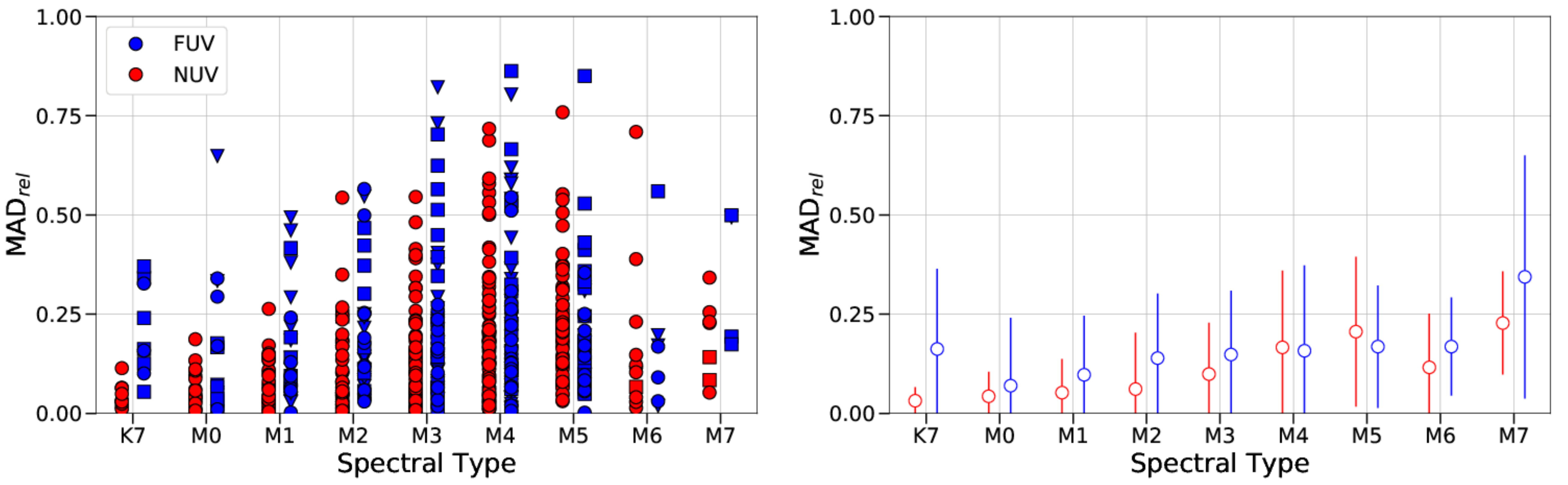}}
\subfigure{\includegraphics[width = 6.5in]{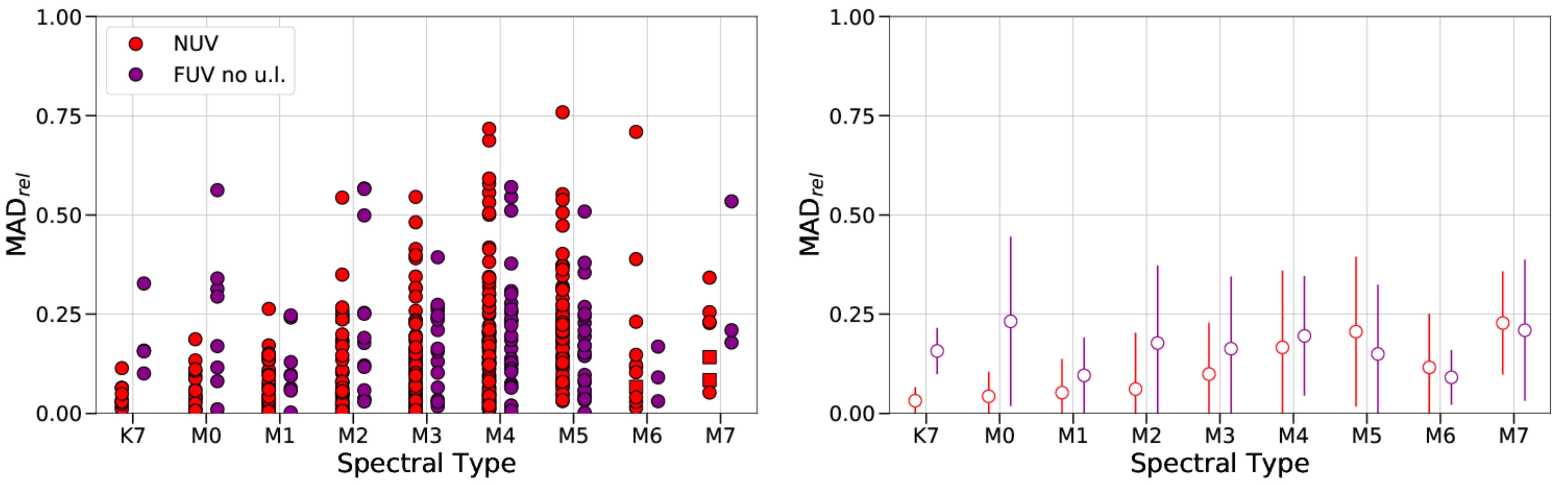}}
\caption{The relative median absolute deviation (MAD$_{rel}$) plotted against spectral type - \textit{Top}: NUV band (red) and FUV band with upper limits (blue). \textit{Bottom}: NUV band (red) and FUV band without upper limits (purple). \textit{Left:} Triangles are used to indicate stars that have only upper limits. Squares are used for stars that have at least one upper limit. Dots represent stars with only detections. \textit{Right:} The open circles are the median spectral type values for each band with the interquartile range plotted.}
\label{fig:nuvcvvsspt}
\end{figure*}

For stars with multiple observations, the median MAD$_{rel}$ in the NUV band is 11\%, with the largest observed change at 76\%. Within each spectral type bin, there is a wide range of variability, with the span increasing with later spectral type (Figure ~\ref{fig:nuvcvvsspt}). M0s have a median variation of 4\% and M4s have a median variation of 16\%. The K7, M6, and M7 bins are each represented by less than a dozen stars. The remaining bins each have between 20 and 85 stars (Figure~\ref{fig:spthist}).

Fifteen percent of the stars with multiple NUV detections are young ($\leq$ 300 Myr). Young stars are known to have higher levels of UV emission in the\GALEX bandpasses relative to older stars \citep{2011ApJ...727....6S,2013MNRAS.431.2063S, 2014AJ....148...64S}. We observe no significant difference between the median variability for each spectral type between the old and young populations.

\begin{figure*}[ht!]
\centering
\includegraphics[width = 6.5in]{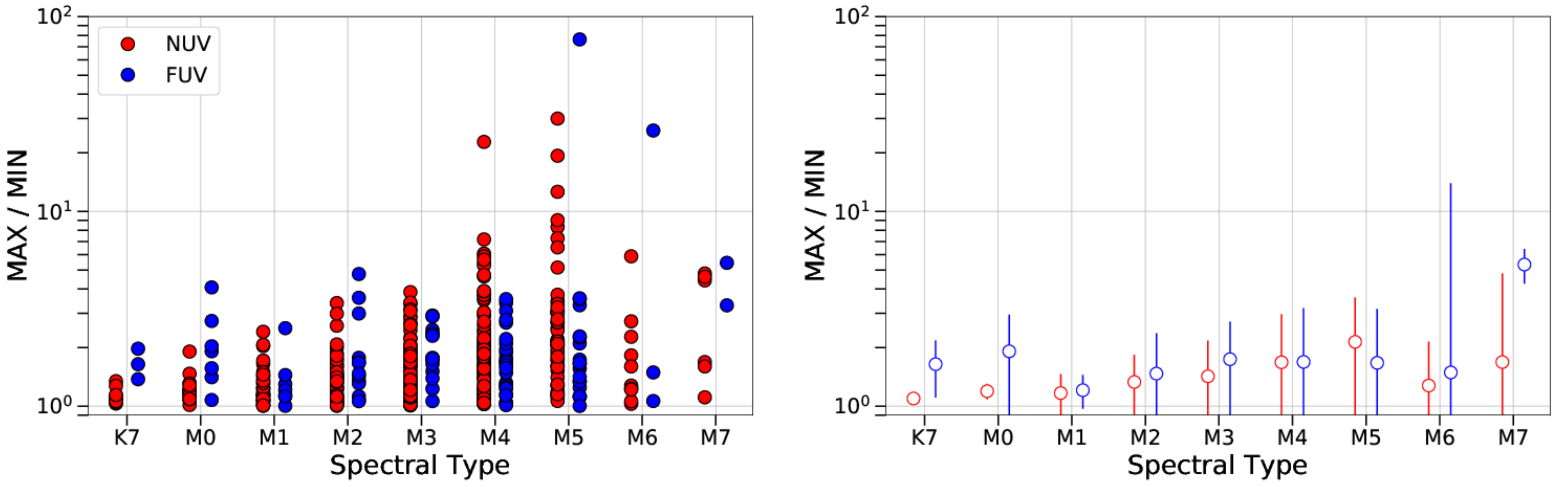}
\caption{The ratio of the maximum flux density to the minimum flux density (MAX$/$MIN) plotted against spectral type for the NUV (red) band and FUV (blue) band with upper limits. \textit{Left}: Dots represent individual stars with MAX/MIN measurements derived from detections. Triangles represent stars where the minimum flux density is estimated with an upper limit. \textit{Right}: Median MAX$/$MIN value for each spectral type plotted with the interquartile range. The ratio for each star is reported in Table~\ref{tbl:starinfo}.}
\label{fig:min_max}
\end{figure*}

The ratio of maximum flux density to minimum flux density (MAX/MIN) has a larger spread and amplitude at later spectral types in the NUV band (Figure~\ref{fig:min_max}). The four stars with a MAX/MIN ratio $>$10 are GJ1167A, GJ3856, LHS3776, and LHS5226. The largest MAX/MIN value was 76.3,  produced by GJ1167A. In the case of these stars, there are distinct flares in the light curves. This is indicated by sharp changes over the course of hours or large deviations from the apparent baseline of each star.

\subsection{FUV Variability}
\label{sec:fuvvar}

\begin{figure}[h!]
\centering
\includegraphics[width = 3in]{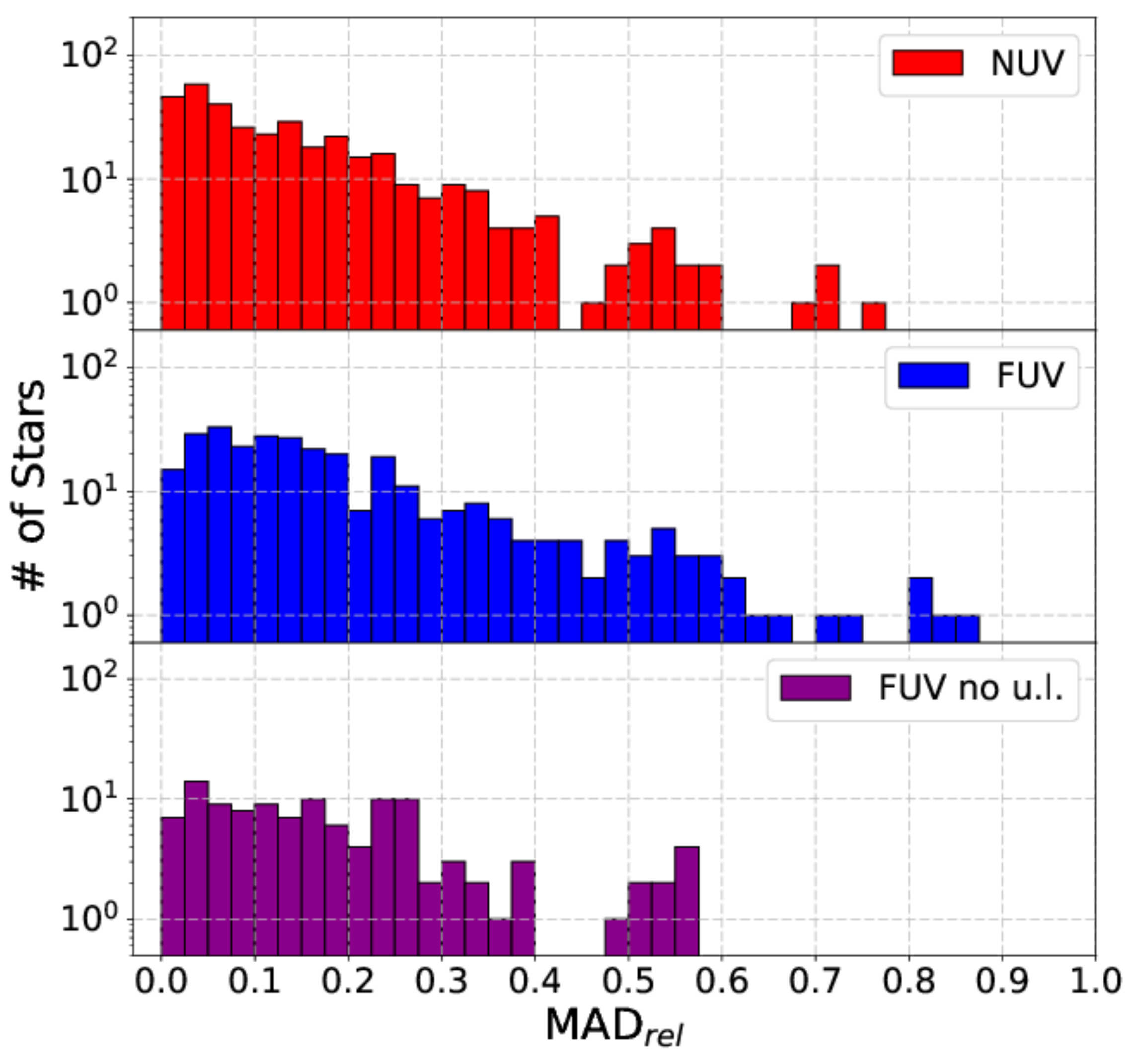}
\caption{Histogram of the relative median absolute deviation (MAD$_{rel}$) for the NUV and FUV bands. FUV is shown with and without upper limits}
\label{fig:mad}
\end{figure}

\begin{figure*}[h!]
\centering
\includegraphics[width = 6.5in]{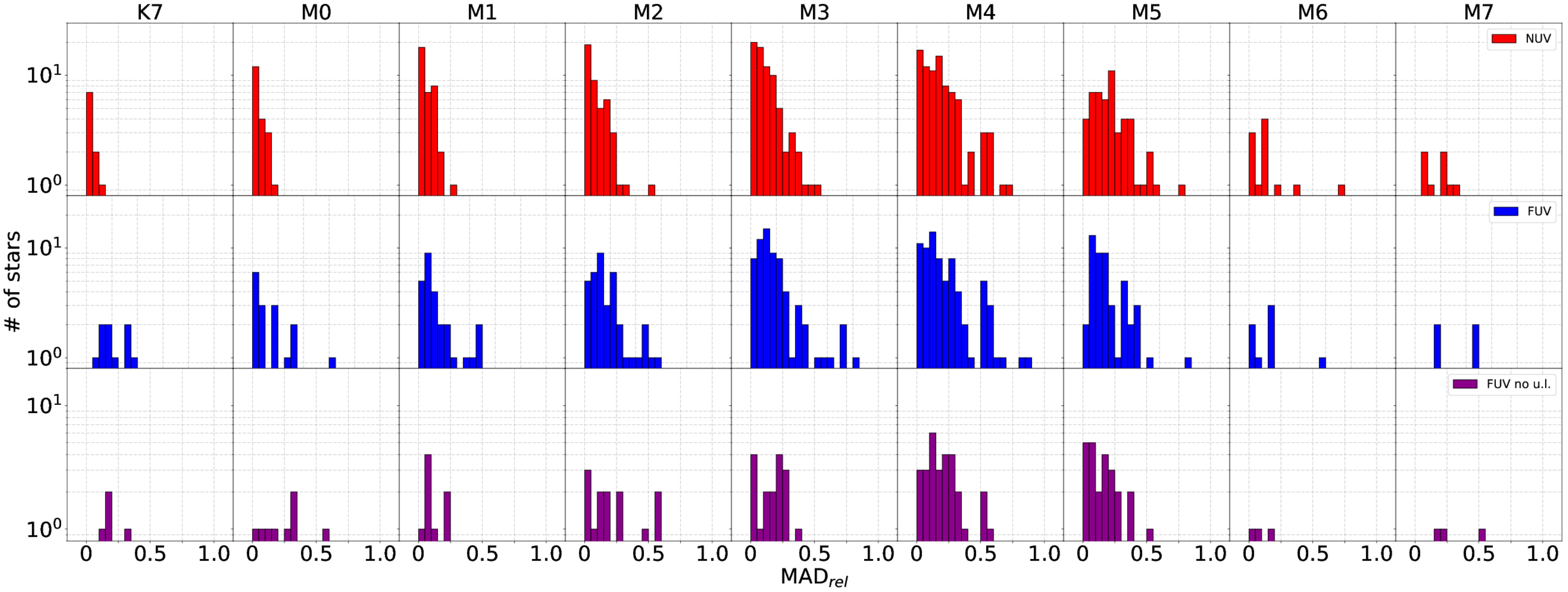}
\caption{Histogram of the relative median absolute deviation (MAD$_{rel}$) for each spectral type. \textit{Top}: NUV band. \textit{Middle}: FUV band with upper limits. \textit{Bottom}: FUV band without upper limits.}
\label{fig:mad_vs_spt}
\end{figure*}

We performed the same variability analysis for the FUV observations as we did for the NUV twice, with and without upper limits because 59\% of the observations were non-detections. The FUV variability spans from 0.1\% to 88\% when the upper limits are included. The median activity in the FUV bandpass is 16\%, which is slightly higher than the NUV value. The distribution of FUV variability without upper limits falls completely within the NUV distribution, although the median activity is higher (Figure~\ref{fig:mad}).
The maximum range of MAD$_{rel}$ goes down to 57\% when the upper limits are removed. In the latter case, the data may be biased to only the brightest and perhaps most active FUV emitters. There remains no significant difference between old and young stars for the overall median levels of variability across spectral types. There is a slight trend between increased variability and later spectral type when upper limits are included, but no clear trend between variability and spectral type when the upper limits are excluded (Figure~\ref{fig:nuvcvvsspt}). The spread in variability between the NUV band and FUV band (with and without upper limits) for each spectral is not significantly different (Figure~\ref{fig:mad_vs_spt}) .

The ratio of maximum flux density to minimum flux density (MAX/MIN) has a larger spread and amplitude at later spectral types in the FUV band (Figure~\ref{fig:min_max}). The four stars that have a MAX/MIN ratio $>$10 are GJ1167A, LHS3776, LHS5094, and UV Cet. The FUV light curves of these stars also show evidence of flaring based on sharp hourly changes and/or large deviations form an apparent baseline. LHS5226 and  GJ3856 do not have detections or upper limits in the FUV despite having relatively high NUV MAX/MIN values. LHS3776 has the largest MAX/MIN in the FUV with a value of 30.0, which is higher than its NUV MAX/MIN value of 24.7. Both LHS5094 and UV Cet have MAX/MIN values in the NUV but they have smaller values, 7.3 and 1.2 respectively.

\begin{deluxetable}{ccccccc}
\tablecaption{\label{tbl:activtysummary}Median MAD$_{rel}$ for each spectral subclass.}
\tabletypesize{\small}
\tablewidth{0pt}
\tablehead{
\colhead{SpT} & \colhead{NUV}  &  \colhead{NUV}  & \colhead{FUV} & \colhead{FUV} & \colhead{FUV} & \colhead{FUV} \\

\colhead{} & \colhead{w/ u.l.\tablenotemark{a}}  &  \colhead{w/ u.l.}  & \colhead{w/ u.l.} & \colhead{w/ u.l.} & \colhead{w/o u.l.\tablenotemark{b}} & \colhead{w/o u.l.} \\

\colhead{} & \colhead{\# stars}  &  \colhead{MAD$_{rel}$}  & \colhead{\# stars} & \colhead{MAD$_{rel}$} & \colhead{\# stars} & \colhead{MAD$_{rel}$}
}
\startdata
K7     & 10       &  0.03 $\pm$ 0.03  &  9  & 0.16 $\pm$ 0.20 & 4  & 0.16 $\pm$ 0.06 \\    
M0     & 20       &  0.04 $\pm$ 0.06  & 16  & 0.07 $\pm$ 0.17 & 8  & 0.23 $\pm$ 0.21 \\ 
M1     & 36       &  0.05 $\pm$ 0.08  & 27  & 0.10 $\pm$ 0.15 & 8  & 0.10 $\pm$ 0.10 \\
M2     & 45       &  0.06 $\pm$ 0.14  & 38  & 0.14 $\pm$ 0.16 & 13 & 0.18 $\pm$ 0.19 \\
M3     & 75       &  0.10 $\pm$ 0.13  & 68  & 0.15 $\pm$ 0.16 & 17 & 0.16 $\pm$ 0.18 \\
M4     & 82       &  0.17 $\pm$ 0.20  & 75  & 0.16 $\pm$ 0.22 & 29 & 0.20 $\pm$ 0.15 \\
M5     & 52       &  0.21 $\pm$ 0.19  & 49  & 0.17 $\pm$ 0.15 & 24 & 0.15 $\pm$ 0.17\\
M6     & 11       &  0.12 $\pm$ 0.14  & 7   & 0.17 $\pm$ 0.12 & 3  & 0.09 $\pm$ 0.07\\
M7     & 7        &  0.23 $\pm$ 0.13  & 4   & 0.34 $\pm$ 0.31 & 3  & 0.21 $\pm$ 0.18\\ 
\hline
Overall& 357      & 0.11    & 303 &  0.15 & 114 & 0.16 \\
\enddata
\vspace{-.5cm}
\tablenotetext{a}{w/ u.l. - with upper limits}
\tablenotetext{b}{w/o u.l. - without upper limits}
\tablecomments{The errors on the MAD$_{rel}$ quantity are the interquartile range. For each bandpass, the overall number of stars does not add up to sum of the stars in each spectral type column because 18 of the stars are identified as M dwarfs, but do not have a published subclass.}
\end{deluxetable}


\begin{figure}[h!]
\centering
\includegraphics[width = 3in]{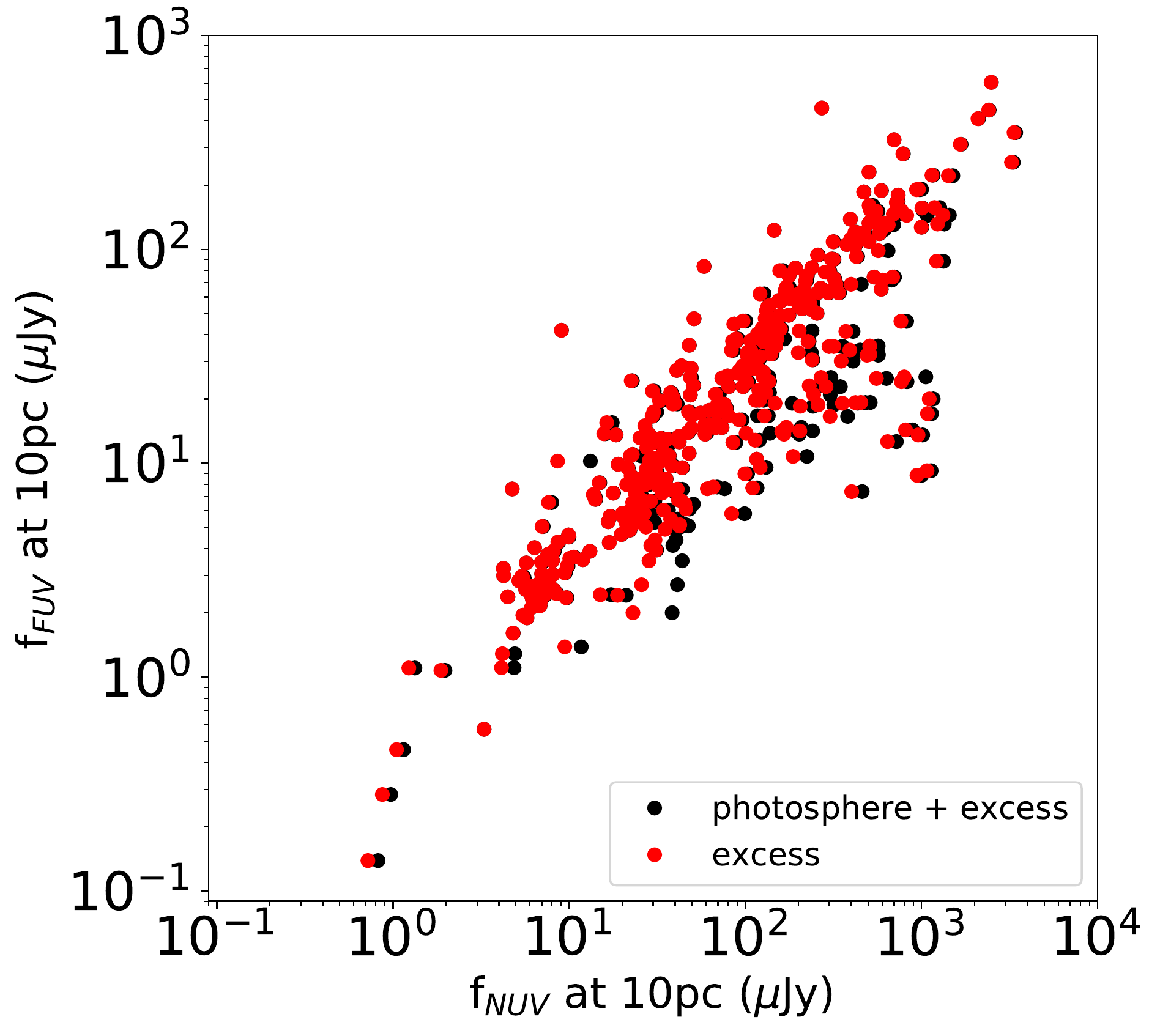}
\caption{FUV flux density ($f_{FUV}$) at 10 pc vs. NUV flux density at 10 pc ($f_{NUV}$). The black dots are the original data and the red dots are the excess flux density.}
\label{fig:fuvvsnuv_phot}
\end{figure}


\section{\GALEX FUV and NUV Correlations and Ratios}
\label{sec:analysis}

Correlations between activity indicators are useful in understanding the physics of stellar atmospheres, as well as providing tools with which to predict one activity diagnostic with the other when it is not observable or detectable. \GALEX\ recorded 393 simultaneous NUV and FUV observations for 145 stars, providing a unique data set with which to search for correlations between the two activity indicators. Detections are considered simultaneous by having the same\GALEX reported time, which is precise down to the second. There are significantly fewer FUV detections than NUV detections because: (1) M dwarfs are typically less luminous in the FUV and (2) the\GALEX FUV detector failed six years into the mission, while the NUV band continued operating for another four years.

We scaled the NUV and FUV detections to 10 pc and subtracted each observation by the photospheric contribution as done by \citet{2014AJ....148...64S} to determine the ``excess'' emission from the stellar upper atmosphere (i.e., the chromosphere, transition region and corona).  This is achieved by using the PHOENIX photospheric models \citep{0004-637X-483-1-390, 0004-637X-618-2-926} for a given stellar effective temperature and age. 

\begin{figure*}[h!]
\centering
\includegraphics[width = 5in]{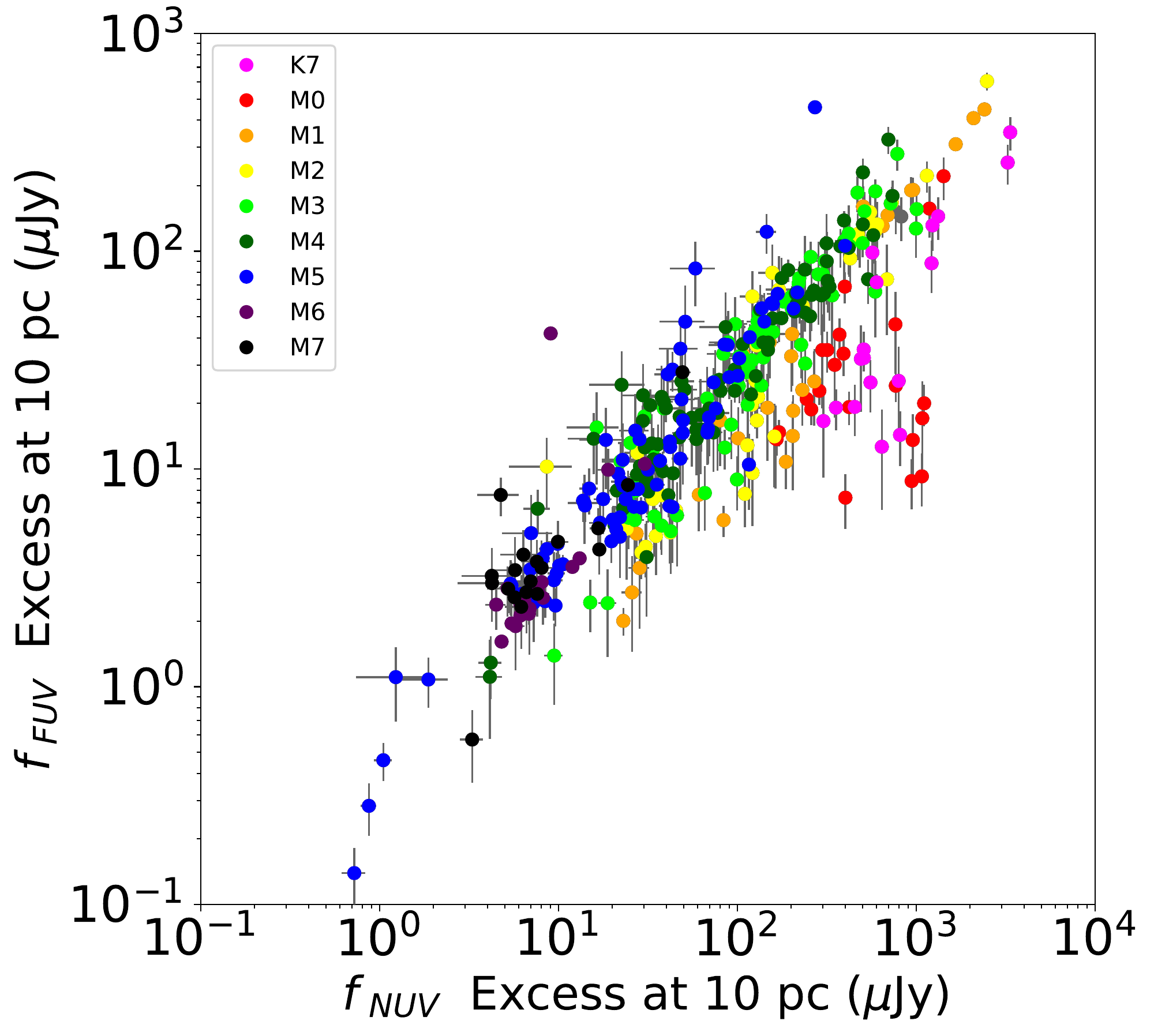}
\caption{Simultaneously observed FUV excess flux density (f$_{FUV}$) vs. NUV excess flux density (f$_{NUV}$)  with colors differentiating spectral types. Coefficients for the equations of the best fit line of each spectral type are listed in Table~\ref{tbl:avg_sim_flux}. }
\label{fig:fuvvsnuv_spt}
\end{figure*}

We converted the published spectral type to effective temperature by interpolating the relationships measured by \cite{1999ApJ...525..466L} and \cite{1982S-K} and compiled by \citet{2007AJ....134.2340K}. Ages were adopted from the literature with the references listed in Table~\ref{tbl:starinfo}. If there is no reported age for a star, we assume it to be the average age of the field, 5 Gyr. The differences are usually negligible between the observed and excess flux densities as shown in Figure~\ref{fig:fuvvsnuv_phot}. On average, the fraction of the model photospheric emission of the observed emission is 4\% and 7$\times$10$^{-4}$\% in the NUV and FUV bands, respectively. As discussed earlier, previously published correlations rely on data sets taken at different times, which introduces excess scatter due to varying activity levels between observations. In Figure~\ref{fig:fuvvsnuv_spt}, we plot all the NUV excess flux densities with simultaneous FUV detections scaled to 10 pc ($f_{FUV}$, $f_{NUV}$). A least squares fit to the excess flux densities  produces the following correlation: 
\begin{equation}
log(f_{FUV}) = A~log(f_{NUV}) + B
\label{eq:log_flux}
\end{equation}
where constants A and B are 0.75 $\pm$ 0.04 and -0.14 $\pm$ 0.07, respectively. 

\begin{figure*}[h!]
\centering
\makebox[\textwidth][c]{\includegraphics[width = 6in]{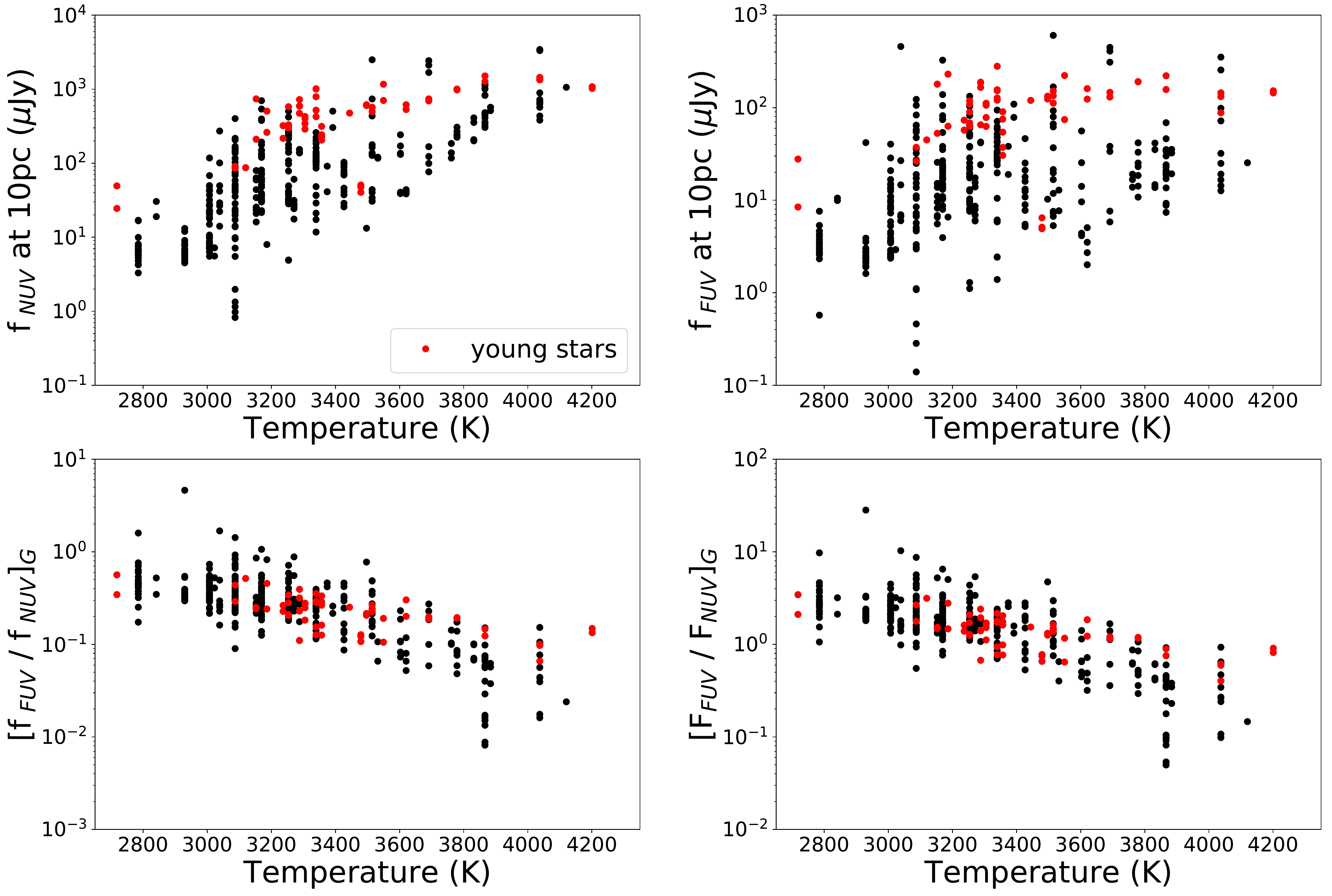}}
\caption{GALEX observed flux densities ($f_{\nu}$) scaled to 10 pc for the NUV (top, left) and FUV (top, right) band plotted against effective temperature. The fraction of FUV to NUV observed flux density (bottom, left) and fraction of integrated band flux ($F_{\nu}$) (bottom, right) plotted against effective temperature. Young stars are have ages less than 300 Myr old and labeled in red.}
\label{fig:fluxtemp}
\end{figure*}

The range of excess flux density covers three orders of magnitude (Table~\ref{tbl:avg_sim_flux}, Figure~\ref{fig:fuvvsnuv_spt}) with the M3 -- M5 population spanning the entire range of UV emission. K7 -- M1 stars have the highest levels of NUV and FUV excess flux densities, and deviate the most from the collection of stars. Figure~\ref{fig:fluxtemp} (top) shows the scaled NUV and FUV observed flux densities plotted against effective temperature. There is an increase of flux density with effective temperature in both\GALEX bands with young ($\leq$300 Myr) stars emitting more emission than older stars. These trends are consistent with the work of \cite{2014AJ....148...64S} and \cite{2015ApJ...798...41A}, who used averaged\GALEX flux densities. Despite later M stars showing more variability (in the NUV), they emit less excess UV emission in both\GALEX bands.

\begin{figure*}[h!]
\centering
\makebox[\textwidth][c]{\includegraphics[width = 6in]{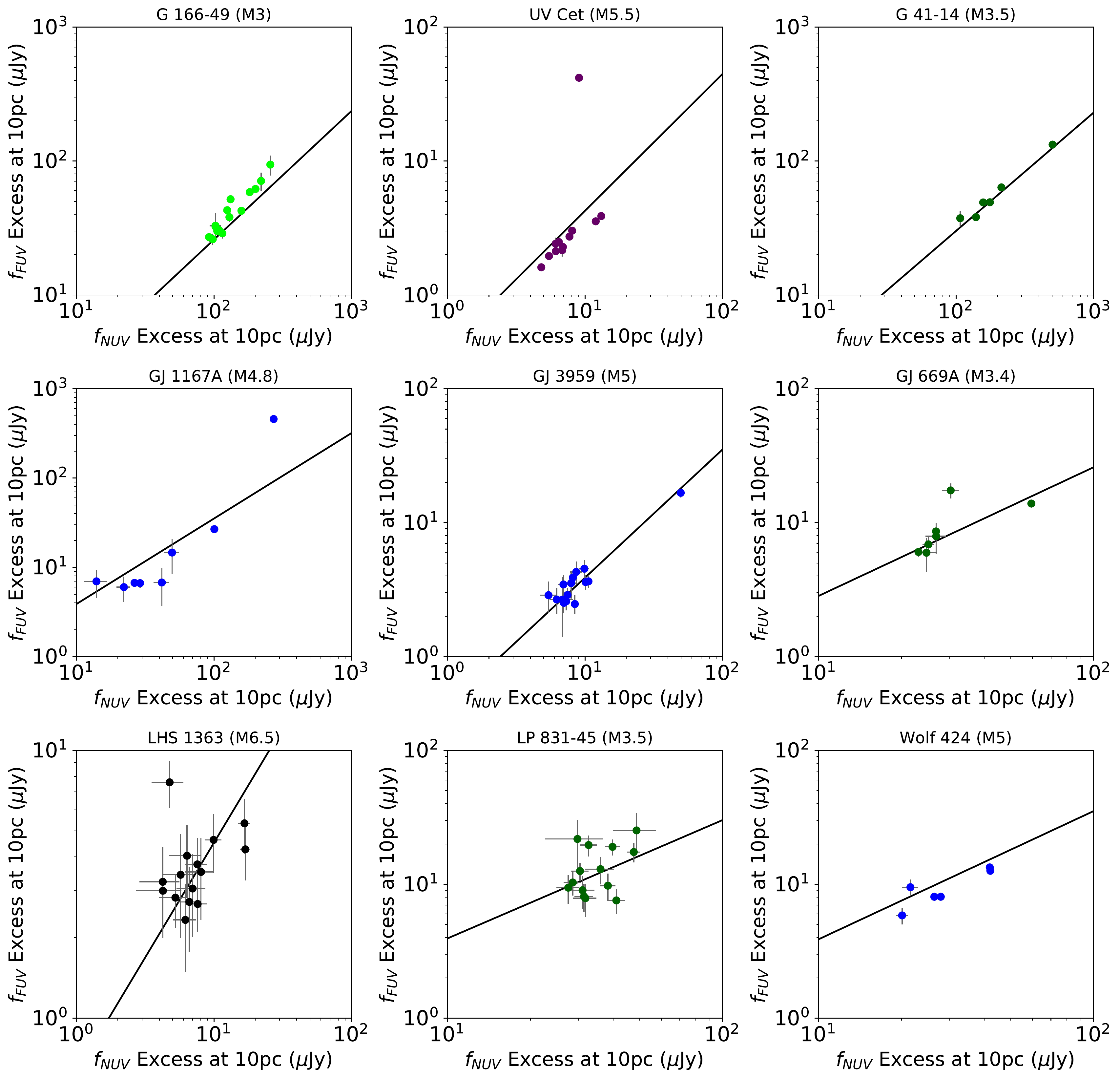}}
\caption{Plots of FUV flux density ($f_{NUV}$) vs NUV flux density ($f_{FUV}$) for individual stars. The black line in every figure is the best fit line of the fluxes for the corresponding spectral type using the coefficients in Table \ref{tbl:avg_sim_flux}. Note: the x and y axis scales are different for most stars.}
\label{fig:indv}
\end{figure*}

Figure~\ref{fig:fluxtemp} (bottom) shows the ratio of the observed FUV to NUV fluxes as a function of effective temperature. The ratio of the integrated fluxes for the \GALEX FUV and NUV bands are denoted by [F$_{FUV}$/F$_{NUV}$]$_{G}$ and the ratio of the flux densities are denoted by  [f$_{FUV}$/f$_{NUV}$]$_{G}$. Note that the FUV band does not include the Lyman-$\alpha$ line where M dwarfs emit a significant portion of energy \citep{2013ApJ...766...69L}. Although cooler stars produce less UV emission overall, both integrated and flux density ratios decrease at higher effective temperatures (Figure~\ref{fig:fluxtemp}; bottom). This trend has also been seen in work by \citet{2016ApJ...820...89F}. The average\GALEX\ flux density ratios for G types stars is 0.01 \citep{2013ApJ...766....9S}, whereas the average flux density ratio for our sample is 31 times higher with a ratio of 0.31.

\citet{2006A&A...458..921W} analyzed time-tagged GALEX data for four M dwarfs from a guest investigator program that monitored each star for 20 to 30 minutes. Their data shows distinct flares in the NUV and FUV. During these flares, the [f$_{FUV}$/f$_{NUV}$]$_{G}$ ratio increases from 0.5 to 13 on the timescale of minutes with clear flare onset occurring at about [f$_{FUV}$/f$_{NUV}$]$_{G}$=1. In our data set, [f$_{FUV}$/f$_{NUV}$]$_{G}$ spans from  0.008 to 4.6 and the 5 [f$_{FUV}$/f$_{NUV}$]$_{G}$ ratios that are $>$1 are likely due to flaring.

\begin{deluxetable}{cccccc}
\tabletypesize{\small}
\tablewidth{0pt}
\tablecaption{\label{tbl:avg_sim_flux} \hspace{.75in} Median Excess Flux Density values at 10pc \newline and Correlations by Spectral Type}
\tablehead{
\colhead{SpT} & \colhead{$f_{NUVe}$\tablenotemark{a}} & \colhead{$f_{FUVe}$} & \colhead{$f_{FUVe}$/$f_{NUVe}$} & \colhead{A\tablenotemark{b}} & \colhead{B\tablenotemark{b}}\\
\colhead{} & \colhead{($\mu$Jy)} & \colhead{($\mu$Jy)} & \colhead{} & \colhead{} & \colhead{}}
\startdata
K7 & 600.4 $\pm$ 890.3 & 32.3 $\pm$ 92.2 & 0.06 $\pm$ 0.04  & 1.24 $\pm$ 1.54  & -1.91 $\pm$ 4.44 \\
M0 & 398.6 $\pm$ 383.1 & 20.9 $\pm$ 49.3 & 0.08 $\pm$ 0.05  & 0.34 $\pm$ 1.12  &  0.50 $\pm$ 3.00 \\
M1 & 202.3 $\pm$ 625.2 & 25.2 $\pm$ 122.7 & 0.18 $\pm$ 0.07  & 1.13 $\pm$ 0.27  & -1.14 $\pm$ 0.63 \\
M2 & 127.9 $\pm$ 458.5 & 21.4 $\pm$ 108.1 & 0.22 $\pm$ 0.19 & 0.96 $\pm$ 0.23  & -0.59 $\pm$ 0.49 \\
M3 & 107.7 $\pm$ 215.8 & 31.0 $\pm$ 52.1 & 0.28 $\pm$ 0.13  & 0.96 $\pm$ 0.20  & -0.51 $\pm$ 0.41 \\
M4 & 71.7  $\pm$160.4 & 22.4 $\pm$ 50.7 & 0.28 $\pm$ 0.16  & 0.88 $\pm$ 0.16  & -0.29 $\pm$ 0.31 \\
M5 & 25.1  $\pm$ 65.5  & 8.1 $\pm$ 52.0 & 0.36 $\pm$ 0.24 & 0.96 $\pm$ 0.10  & -0.37 $\pm$ 0.14 \\
M6 & 6.9   $\pm$ 6.2   & 2.5  $\pm$ 9.2 & 0.35 $\pm$ 0.98  & 1.02 $\pm$ 0.63  & -0.39 $\pm$ 0.57 \\
M7 & 6.6 $\pm$ 10.5  & 3.4 $\pm$ 5.6 & 0.45 $\pm$ 0.29   & 0.86 $\pm$ 0.61  & -0.20 $\pm$ 0.55 \\
\enddata
\vspace{-0.5cm}
\tablecomments{\label{tbl:avg_sim_flux}For the first two columns, the errors are the standard deviation of the stars within the spectral type bin.}
\tablenotetext{a}{$f_{FUVe}$,$f_{NUVe}$ - excess flux density from layers above the photosphere.}
\tablenotetext{b}{A and B are the coefficients from Equation~\ref{eq:log_flux}}
\end{deluxetable}

Nine stars with at least five simultaneous NUV and FUV detections were investigated individually to see how well their observations match the best fit line found using all simultaneous data of the same spectral type. The FUV and NUV flux densities of these stars are plotted in Figure~\ref{fig:indv}. Five of the stars show comparable correlations to their respective spectral type, but four have observations which appear as a scatter plot. For three of the nine stars, we observe an order of magnitude spread in emission in one or both of the\GALEX bands.
UV Cet (M5.5) and G166-49 (M3) have the highest signal to noise and time resolution of the nine targets (Figure~\ref{fig:lightcurves}). In Figure~\ref{fig:indv} (top), all of UV Cet's data points follow near the best fit line for M5s, except for one observation when only the FUV emission increases by an order of magnitude. This causes [f$_{FUV}$/f$_{NUV}$]$_{G}$ to jump from a baseline of $\sim$0.5 to 4.6. Such a sharp increase in FUV activity over a few hours is most likely due to a flare.


\section{Conclusions}

We analyzed the 377 low-mass stars within 30 pc with multiple photometric observations in the\GALEX archive to characterize the stellar variability with the NUV and FUV bandpasses. The timing between the observations for each target ranges from minutes to years with a median time between observations of about a year. There were simultaneous NUV and FUV observations for 145 of the stars, which we used to measure the correlation between NUV and FUV emission for each spectral type and the variations in the\GALEX FUV to NUV ratio. 

The summary of our findings includes:

\textbf{\textit{UV Variability:}} The median variation as measured by the relative median absolute deviation (MAD$_{rel}$) in the NUV flux density is 11\% for the entire sample, and 16\% for the FUV flux density. The median level of stellar variability in the NUV bandpass increases with later spectral type, from  4\% for M0s to 21\% for M5s. When upper limits are included, there is a slight trend between increased variability and later spectral type for the FUV band. There is no clear trend  with spectral type when upper limits are not used. The median variability is 16\% for M0s and 15\% for M5s.The variability in the NUV and FUV is not significantly different for the young stars ($\leq$300 Myr) compared to the old stars in our sample. The ratio of maximum flux density to minimum flux density has a larger spread at later spectral types for both the NUV and FUV bands.

\textbf{\textit{\GALEX FUV and NUV Correlations:}}  
The excess FUV and NUV flux densities (i.e.~with effective temperature dependent photospheric emission substracted) are correlated for stars with spectral types between M1 and M6. In these cases, the NUV emission can act as a proxy for the FUV when there are no FUV observations or detections. K7, M0 and M7 stars show no clear correlation in our data set, possibly due to the low numbers of stars in each bin and their narrower range in emission levels.  

\textbf{\textit{\GALEX [f$_{FUV}$/f$_{NUV}$]$_{G}$ Ratio:}} The average of FUV to NUV flux density ratio of our low-mass stars are 31 times higher than the average for G-stars measured with \GALEX, bolstering the need to measure such ratios for the low-mass stars that host HZ planets with atmospheres in which observers will seek oxygen. \cite{2014E&PSL.385...22T} calculated that under these UV conditions, O$_2$ and O$_3$ could be 2--3 orders of magnitude greater than in the atmospheres of HZ planets around Sun-like stars producing a potential false-positive detection of a biosignature \citep{2015ApJ...812..137H}. In a few cases, likely during a flare, we see significant deviation of the\GALEX [f$_{FUV}$/f$_{NUV}$]$_{G}$ ratio from the norm for a given star reaching levels $>$1. We also observe that on average,\GALEX [f$_{FUV}$/f$_{NUV}$]$_{G}$ and [F$_{FUV}$/F$_{NUV}$]$_{G}$ for each observation increases for later spectral types with. For M0s f$_{FUV}$/f$_{NUV}$ has a median value of 0.08 and M4s has a median value of 0.28.   

These results characterize the UV behavior for the largest set of low-mass stars to date. The statistical FUV and NUV variability levels, correlations and ratios will aid in our understanding of the high-energy radiation environment of exoplanets. Future studies with greater time resolution and temporal coverage should reliably distinguish between baseline emission and modulation due to rotation, flaring and stellar activity cycles, especially valuable for a particular exoplanetary system of interest. This will require building dedicated UV telescopes that can observe a sizable population of low-mass stars and monitor choice planet hosts for weeks at a time. Understanding such stellar activity will become especially important when missions like the James Webb Space Telescope begin to search for biosignatures in the spectra of HZ planets around low-mass stars. 

\section{Acknowledgements}
The authors would like to thank S. Flemming and T. S. Barman for useful discussion. B.E.M acknowledges the support of NSF grant AST-1461200 to Northern Arizona University and the National Institute of General Medical Sciences of the National Institutes of Health under award number R25GM055052 awarded to T. Hasson. The content is solely the responsibility of the authors and does not necessarily represent the official views of the National Institutes of Health. E.S. appreciates support from NASA/Habitable Worlds grant NNX16AB62G.



\begin{thebibliography}{}
\expandafter\ifx\csname natexlab\endcsname\relax\def\natexlab#1{#1}\fi

\bibitem[{{Alonso-Floriano} {et~al.}(2015){Alonso-Floriano}, {Morales},
  {Caballero}, {Montes}, {Klutsch}, {Mundt}, {Cort{\'e}s-Contreras}, {Ribas},
  {Reiners}, {Amado}, {Quirrenbach}, \& {Jeffers}}]{2015A&A...577A.128A}
{Alonso-Floriano}, F.~J., {Morales}, J.~C., {Caballero}, J.~A., {et~al.} 2015,
  \aap, 577, A128

\bibitem[{{Anglada-Escud{\'e}} {et~al.}(2016){Anglada-Escud{\'e}}, {Amado},
  {Barnes}, {Berdi{\~n}as}, {Butler}, {Coleman}, {de La Cueva}, {Dreizler},
  {Endl}, {Giesers}, {Jeffers}, {Jenkins}, {Jones}, {Kiraga}, {K{\"u}rster},
  {L{\'o}pez-Gonz{\'a}lez}, {Marvin}, {Morales}, {Morin}, {Nelson}, {Ortiz},
  {Ofir}, {Paardekooper}, {Reiners}, {Rodr{\'{\i}}guez},
  {Rodr{\'{\i}}guez-L{\'o}pez}, {Sarmiento}, {Strachan}, {Tsapras}, {Tuomi}, \&
  {Zechmeister}}]{2016Natur.536..437A}
{Anglada-Escud{\'e}}, G., {Amado}, P.~J., {Barnes}, J., {et~al.} 2016, \nat,
  536, 437

\bibitem[{{Ansdell} {et~al.}(2015){Ansdell}, {Gaidos}, {Mann}, {L{\'e}pine},
  {James}, {Buccino}, {Baranec}, {Law}, {Riddle}, {Mauas}, \&
  {Petrucci}}]{2015ApJ...798...41A}
{Ansdell}, M., {Gaidos}, E., {Mann}, A.~W., {et~al.} 2015, \apj, 798, 41

\bibitem[{{Arney} {et~al.}(2016){Arney}, {Domagal-Goldman}, {Meadows}, {Wolf},
  {Schwieterman}, {Charnay}, {Claire}, {H{\'e}brard}, \& {Trainer}}]{arne16}
{Arney}, G., {Domagal-Goldman}, S.~D., {Meadows}, V.~S., {et~al.} 2016,
  Astrobiology, 16, 873

\bibitem[{{Bochanski} {et~al.}(2010){Bochanski}, {Hawley}, {Covey}, {West},
  {Reid}, {Golimowski}, \& {Ivezi{\'c}}}]{2010AJ....139.2679B}
{Bochanski}, J.~J., {Hawley}, S.~L., {Covey}, K.~R., {et~al.} 2010, \aj, 139,
  2679

\bibitem[{{Dressing} \& {Charbonneau}(2015)}]{2015ApJ...807...45D}
{Dressing}, C.~D., \& {Charbonneau}, D. 2015, \apj, 807, 45

\bibitem[{{France} {et~al.}(2013){France}, {Froning}, {Linsky}, {Roberge},
  {Stocke}, {Tian}, {Bushinsky}, {D{\'e}sert}, {Mauas}, {Vieytes}, \&
  {Walkowicz}}]{2013ApJ...763..149F}
{France}, K., {Froning}, C.~S., {Linsky}, J.~L., {et~al.} 2013, \apj, 763, 149

\bibitem[{{France} {et~al.}(2016){France}, {Parke Loyd}, {Youngblood}, {Brown},
  {Schneider}, {Hawley}, {Froning}, {Linsky}, {Roberge}, {Buccino},
  {Davenport}, {Fontenla}, {Kaltenegger}, {Kowalski}, {Mauas}, {Miguel},
  {Redfield}, {Rugheimer}, {Tian}, {Vieytes}, {Walkowicz}, \&
  {Weisenburger}}]{2016ApJ...820...89F}
{France}, K., {Parke Loyd}, R.~O., {Youngblood}, A., {et~al.} 2016, \apj, 820,
  89

\bibitem[{{Gillon} {et~al.}(2017){Gillon}, {Triaud}, {Demory}, {Jehin}, {Agol},
  {Deck}, {Lederer}, {de Wit}, {Burdanov}, {Ingalls}, {Bolmont}, {Leconte},
  {Raymond}, {Selsis}, {Turbet}, {Barkaoui}, {Burgasser}, {Burleigh}, {Carey},
  {Chaushev}, {Copperwheat}, {Delrez}, {Fernandes}, {Holdsworth}, {Kotze}, {Van
  Grootel}, {Almleaky}, {Benkhaldoun}, {Magain}, \&
  {Queloz}}]{2017Natur.542..456G}
{Gillon}, M., {Triaud}, A.~H.~M.~J., {Demory}, B.-O., {et~al.} 2017, \nat, 542,
  456

\bibitem[{{Gray} {et~al.}(2006){Gray}, {Corbally}, {Garrison}, {McFadden},
  {Bubar}, {McGahee}, {O'Donoghue}, \& {Knox}}]{2006AJ....132..161G}
{Gray}, R.~O., {Corbally}, C.~J., {Garrison}, R.~F., {et~al.} 2006, \aj, 132,
  161

\bibitem[{{Gray} {et~al.}(2003){Gray}, {Corbally}, {Garrison}, {McFadden}, \&
  {Robinson}}]{2003AJ....126.2048G}
{Gray}, R.~O., {Corbally}, C.~J., {Garrison}, R.~F., {McFadden}, M.~T., \&
  {Robinson}, P.~E. 2003, \aj, 126, 2048

\bibitem[{{Harman} {et~al.}(2015){Harman}, {Schwieterman}, {Schottelkotte}, \&
  {Kasting}}]{2015ApJ...812..137H}
{Harman}, C.~E., {Schwieterman}, E.~W., {Schottelkotte}, J.~C., \& {Kasting},
  J.~F. 2015, \apj, 812, 137

\bibitem[{{Harris} {et~al.}(1988){Harris}, {Gry}, {Bohlin}, {Blades}, \&
  {Holm}}]{1988ESASP.281b.365H}
{Harris}, A.~W., {Gry}, C., {Bohlin}, R.~C., {Blades}, J.~C., \& {Holm}, A.~V.
  1988, in ESA Special Publication, Vol. 281, ESA Special Publication

\bibitem[{Hauschildt {et~al.}(1997)Hauschildt, Baron, \&
  Allard}]{0004-637X-483-1-390}
Hauschildt, P.~H., Baron, E., \& Allard, F. 1997, The Astrophysical Journal,
  483, 390

\bibitem[{{Hawley} {et~al.}(1996){Hawley}, {Gizis}, \&
  {Reid}}]{1996AJ....112.2799H}
{Hawley}, S.~L., {Gizis}, J.~E., \& {Reid}, I.~N. 1996, \aj, 112, 2799

\bibitem[{{Jenkins} {et~al.}(2009){Jenkins}, {Ramsey}, {Jones}, {Pavlenko},
  {Gallardo}, {Barnes}, \& {Pinfield}}]{2009ApJ...704..975J}
{Jenkins}, J.~S., {Ramsey}, L.~W., {Jones}, H.~R.~A., {et~al.} 2009, \apj, 704,
  975

\bibitem[{{King} {et~al.}(2003){King}, {Villarreal}, {Soderblom}, {Gulliver},
  \& {Adelman}}]{2003AJ....125.1980K}
{King}, J.~R., {Villarreal}, A.~R., {Soderblom}, D.~R., {Gulliver}, A.~F., \&
  {Adelman}, S.~J. 2003, \aj, 125, 1980

\bibitem[{{Kraus} \& {Hillenbrand}(2007)}]{2007AJ....134.2340K}
{Kraus}, A.~L., \& {Hillenbrand}, L.~A. 2007, \aj, 134, 2340

\bibitem[{{Kraus} {et~al.}(2014){Kraus}, {Shkolnik}, {Allers}, \&
  {Liu}}]{2014AJ....147..146K}
{Kraus}, A.~L., {Shkolnik}, E.~L., {Allers}, K.~N., \& {Liu}, M.~C. 2014, \aj,
  147, 146

\bibitem[{{Kretzschmar} {et~al.}(2009){Kretzschmar}, {Dudok de Wit},
  {Lilensten}, {Hochedez}, {Aboudarham}, {Amblard}, {Auch{\`e}re}, \&
  {Moussaoui}}]{2009AcGeo..57...42K}
{Kretzschmar}, M., {Dudok de Wit}, T., {Lilensten}, J., {et~al.} 2009, Acta
  Geophysica, 57, 42

\bibitem[{{L{\'e}pine} {et~al.}(2013){L{\'e}pine}, {Hilton}, {Mann}, {Wilde},
  {Rojas-Ayala}, {Cruz}, \& {Gaidos}}]{2013AJ....145..102L}
{L{\'e}pine}, S., {Hilton}, E.~J., {Mann}, A.~W., {et~al.} 2013, \aj, 145, 102

\bibitem[{{Linsky} {et~al.}(2013){Linsky}, {France}, \&
  {Ayres}}]{2013ApJ...766...69L}
{Linsky}, J.~L., {France}, K., \& {Ayres}, T. 2013, \apj, 766, 69

\bibitem[{{Loyd} \& {France}(2014)}]{2014ApJS..211....9L}
{Loyd}, R.~O.~P., \& {France}, K. 2014, \apjs, 211, 9

\bibitem[{{Luger} \& {Barnes}(2015)}]{2015AsBio..15..119L}
{Luger}, R., \& {Barnes}, R. 2015, Astrobiology, 15, 119

\bibitem[{{Luger} {et~al.}(2015){Luger}, {Barnes}, {Lopez}, {Fortney},
  {Jackson}, \& {Meadows}}]{2015AsBio..15...57L}
{Luger}, R., {Barnes}, R., {Lopez}, E., {et~al.} 2015, Astrobiology, 15, 57

\bibitem[{{Luhman}(1999)}]{1999ApJ...525..466L}
{Luhman}, K.~L. 1999, \apj, 525, 466

\bibitem[{{Mann} {et~al.}(2015){Mann}, {Feiden}, {Gaidos}, {Boyajian}, \& {von
  Braun}}]{2015ApJ...804...64M}
{Mann}, A.~W., {Feiden}, G.~A., {Gaidos}, E., {Boyajian}, T., \& {von Braun},
  K. 2015, \apj, 804, 64

\bibitem[{{McCook} \& {Sion}(1999)}]{1999ApJS..121....1M}
{McCook}, G.~P., \& {Sion}, E.~M. 1999, \apjs, 121, 1

\bibitem[{{Miguel} \& {Kaltenegger}(2014)}]{2014ApJ...780..166M}
{Miguel}, Y., \& {Kaltenegger}, L. 2014, \apj, 780, 166

\bibitem[{{Mitra-Kraev} {et~al.}(2005){Mitra-Kraev}, {Harra}, {G{\"u}del},
  {Audard}, {Branduardi-Raymont}, {Kay}, {Mewe}, {Raassen}, \& {van
  Driel-Gesztelyi}}]{2005A&A...431..679M}
{Mitra-Kraev}, U., {Harra}, L.~K., {G{\"u}del}, M., {et~al.} 2005, \aap, 431,
  679

\bibitem[{{Montes} {et~al.}(2001){Montes}, {L{\'o}pez-Santiago}, {G{\'a}lvez},
  {Fern{\'a}ndez-Figueroa}, {De Castro}, \& {Cornide}}]{2001MNRAS.328...45M}
{Montes}, D., {L{\'o}pez-Santiago}, J., {G{\'a}lvez}, M.~C., {et~al.} 2001,
  \mnras, 328, 45

\bibitem[{{Morrissey} {et~al.}(2005){Morrissey}, {Schiminovich}, {Barlow},
  {Martin}, {Blakkolb}, {Conrow}, {Cooke}, {Erickson}, {Fanson}, {Friedman},
  {Grange}, {Jelinsky}, {Lee}, {Liu}, {Mazer}, {McLean}, {Milliard}, {Randall},
  {Schmitigal}, {Sen}, {Siegmund}, {Surber}, {Vaughan}, {Viton}, {Welsh},
  {Bianchi}, {Byun}, {Donas}, {Forster}, {Heckman}, {Lee}, {Madore}, {Malina},
  {Neff}, {Rich}, {Small}, {Szalay}, \& {Wyder}}]{2005ApJ...619L...7M}
{Morrissey}, P., {Schiminovich}, D., {Barlow}, T.~A., {et~al.} 2005, \apjl,
  619, L7

\bibitem[{{Morrissey} {et~al.}(2007){Morrissey}, {Conrow}, {Barlow}, {Small},
  {Seibert}, {Wyder}, {Budav{\'a}ri}, {Arnouts}, {Friedman}, {Forster},
  {Martin}, {Neff}, {Schiminovich}, {Bianchi}, {Donas}, {Heckman}, {Lee},
  {Madore}, {Milliard}, {Rich}, {Szalay}, {Welsh}, \&
  {Yi}}]{2007ApJS..173..682M}
{Morrissey}, P., {Conrow}, T., {Barlow}, T.~A., {et~al.} 2007, \apjs, 173, 682

\bibitem[{{Newton} {et~al.}(2014){Newton}, {Charbonneau}, {Irwin},
  {Berta-Thompson}, {Rojas-Ayala}, {Covey}, \& {Lloyd}}]{2014AJ....147...20N}
{Newton}, E.~R., {Charbonneau}, D., {Irwin}, J., {et~al.} 2014, \aj, 147, 20

\bibitem[{{Nidever} {et~al.}(2002){Nidever}, {Marcy}, {Butler}, {Fischer}, \&
  {Vogt}}]{2002ApJS..141..503N}
{Nidever}, D.~L., {Marcy}, G.~W., {Butler}, R.~P., {Fischer}, D.~A., \& {Vogt},
  S.~S. 2002, \apjs, 141, 503

\bibitem[{{Perryman} {et~al.}(1997){Perryman}, {Lindegren}, {Kovalevsky},
  {Hoeg}, {Bastian}, {Bernacca}, {Cr{\'e}z{\'e}}, {Donati}, {Grenon},
  {Grewing}, {van Leeuwen}, {van der Marel}, {Mignard}, {Murray}, {Le Poole},
  {Schrijver}, {Turon}, {Arenou}, {Froeschl{\'e}}, \&
  {Petersen}}]{1997A&A...323L..49P}
{Perryman}, M.~A.~C., {Lindegren}, L., {Kovalevsky}, J., {et~al.} 1997, \aap,
  323, L49

\bibitem[{{Rajpurohit} {et~al.}(2013){Rajpurohit}, {Reyl{\'e}}, {Allard},
  {Homeier}, {Schultheis}, {Bessell}, \& {Robin}}]{2013A&A...556A..15R}
{Rajpurohit}, A.~S., {Reyl{\'e}}, C., {Allard}, F., {et~al.} 2013, \aap, 556,
  A15

\bibitem[{{Reid} \& {Cruz}(2002)}]{2002AJ....123.2806R}
{Reid}, I.~N., \& {Cruz}, K.~L. 2002, \aj, 123, 2806

\bibitem[{{Reid} {et~al.}(2007){Reid}, {Cruz}, \&
  {Allen}}]{2007AJ....133.2825R}
{Reid}, I.~N., {Cruz}, K.~L., \& {Allen}, P.~R. 2007, \aj, 133, 2825

\bibitem[{{Reid} {et~al.}(1995){Reid}, {Hawley}, \&
  {Gizis}}]{1995AJ....110.1838R}
{Reid}, I.~N., {Hawley}, S.~L., \& {Gizis}, J.~E. 1995, \aj, 110, 1838

\bibitem[{{Rugheimer} {et~al.}(2015){Rugheimer}, {Kaltenegger}, {Segura},
  {Linsky}, \& {Mohanty}}]{2015ApJ...809...57R}
{Rugheimer}, S., {Kaltenegger}, L., {Segura}, A., {Linsky}, J., \& {Mohanty},
  S. 2015, \apj, 809, 57

\bibitem[{{Schlieder} {et~al.}(2012){Schlieder}, {L{\'e}pine}, \&
  {Simon}}]{2012AJ....143...80S}
{Schlieder}, J.~E., {L{\'e}pine}, S., \& {Simon}, M. 2012, \aj, 143, 80

\bibitem[{{Schmidt-Kaler}(1982)}]{1982S-K}
{Schmidt-Kaler}, T. 1982, "Physical Parameters of the Stars", Vol.~2
  (Springer-Verlag, Berlin)

\bibitem[{{Scholz} {et~al.}(2005){Scholz}, {Meusinger}, \&
  {Jahrei{\ss}}}]{2005A&A...442..211S}
{Scholz}, R.-D., {Meusinger}, H., \& {Jahrei{\ss}}, H. 2005, \aap, 442, 211

\bibitem[{{Segura} {et~al.}(2010){Segura}, {Walkowicz}, {Meadows}, {Kasting},
  \& {Hawley}}]{2010AsBio..10..751S}
{Segura}, A., {Walkowicz}, L.~M., {Meadows}, V., {Kasting}, J., \& {Hawley}, S.
  2010, Astrobiology, 10, 751

\bibitem[{{Shkolnik} {et~al.}(2009){Shkolnik}, {Liu}, \&
  {Reid}}]{2009ApJ...699..649S}
{Shkolnik}, E., {Liu}, M.~C., \& {Reid}, I.~N. 2009, \apj, 699, 649

\bibitem[{{Shkolnik}(2013)}]{2013ApJ...766....9S}
{Shkolnik}, E.~L. 2013, \apj, 766, 9

\bibitem[{{Shkolnik} {et~al.}(2012){Shkolnik}, {Anglada-Escud{\'e}}, {Liu},
  {Bowler}, {Weinberger}, {Boss}, {Reid}, \& {Tamura}}]{2012ApJ...758...56S}
{Shkolnik}, E.~L., {Anglada-Escud{\'e}}, G., {Liu}, M.~C., {et~al.} 2012, \apj,
  758, 56

\bibitem[{{Shkolnik} \& {Barman}(2014)}]{2014AJ....148...64S}
{Shkolnik}, E.~L., \& {Barman}, T.~S. 2014, \aj, 148, 64

\bibitem[{{Shkolnik} {et~al.}(2011){Shkolnik}, {Liu}, {Reid}, {Dupuy}, \&
  {Weinberger}}]{2011ApJ...727....6S}
{Shkolnik}, E.~L., {Liu}, M.~C., {Reid}, I.~N., {Dupuy}, T., \& {Weinberger},
  A.~J. 2011, \apj, 727, 6

\bibitem[{{Shkolnik} {et~al.}(2014){Shkolnik}, {Rolph}, {Peacock}, \&
  {Barman}}]{2014ApJ...796L..20S}
{Shkolnik}, E.~L., {Rolph}, K.~A., {Peacock}, S., \& {Barman}, T.~S. 2014,
  \apjl, 796, L20

\bibitem[{Short \& Hauschildt(2005)}]{0004-637X-618-2-926}
Short, C.~I., \& Hauschildt, P.~H. 2005, The Astrophysical Journal, 618, 926

\bibitem[{{Stelzer} {et~al.}(2013){Stelzer}, {Marino}, {Micela},
  {L{\'o}pez-Santiago}, \& {Liefke}}]{2013MNRAS.431.2063S}
{Stelzer}, B., {Marino}, A., {Micela}, G., {L{\'o}pez-Santiago}, J., \&
  {Liefke}, C. 2013, \mnras, 431, 2063

\bibitem[{{Tian} {et~al.}(2014){Tian}, {France}, {Linsky}, {Mauas}, \&
  {Vieytes}}]{2014E&PSL.385...22T}
{Tian}, F., {France}, K., {Linsky}, J.~L., {Mauas}, P.~J.~D., \& {Vieytes},
  M.~C. 2014, Earth and Planetary Science Letters, 385, 22

\bibitem[{{Torres} {et~al.}(2006){Torres}, {Quast}, {da Silva}, {de La Reza},
  {Melo}, \& {Sterzik}}]{2006A&A...460..695T}
{Torres}, C.~A.~O., {Quast}, G.~R., {da Silva}, L., {et~al.} 2006, \aap, 460,
  695

\bibitem[{{Welsh} {et~al.}(2006){Welsh}, {Wheatley}, {Browne}, {Siegmund},
  {Doyle}, {O'Shea}, {Antonova}, {Forster}, {Seibert}, {Morrissey}, \&
  {Taroyan}}]{2006A&A...458..921W}
{Welsh}, B.~Y., {Wheatley}, J., {Browne}, S.~E., {et~al.} 2006, \aap, 458, 921

\bibitem[{{Wheatley} {et~al.}(2008){Wheatley}, {Welsh}, \&
  {Browne}}]{2008AJ....136..259W}
{Wheatley}, J.~M., {Welsh}, B.~Y., \& {Browne}, S.~E. 2008, \aj, 136, 259

\end{thebibliography}
\end{document}